\documentstyle[graphicx,latexsym,amsfonts,prb,aps
,multicol%
,color%
]{revtex}
\newcommand{\Mmulti}{\begin{multicols}{2}}
\newcommand{\Mmultii}{\end{multicols}}

\newcommand{\DTK}{d^2{\mathbf k}\;}
\newcommand{\DTR}{d^3{\mathbf r}\;}
\newcommand{\Vol} [1]{{\bf #1}}
\newcommand{\VECk}{{\mathbf k}}
\newcommand{\ef}{E_{\scriptscriptstyle F}}
\newcommand{\Mfactor}{.92}
\begin{document}
\draft

\title{
Evolution of Co/Cu multilayer conductivity during growth:\\
ab intio study
}

\author{P.~Zahn$^a$, N.~Papanikolaou$^b$, F.~Erler$^a$, and I.~Mertig$^b$}
\address{ $^a$ Technische Universit\"{a}t Dresden, 
Institut f\"{u}r Theoretische Physik,
D-01062 Dresden, Germany}
\address{ $^b$ Martin-Luther-Universit\"{a}t, 
Halle-Wittenberg, 
Fachbereich Physik, Fachgruppe Theoretische Physik,
D-06099 Halle, Germany}
\date{\today}
\maketitle

\begin{abstract}
We present ab-initio calculations for the in plane conductivity of
Co/Cu multilayer slabs. The electronic structure of the multilayer
slabs is calculated by means of density functional theory within a
screened KKR scheme.
Transport properties are described using the Boltzmann equation in relaxation 
time approximation. We study the change of the conductivity during
growth of the multilayer, 
and we can reproduce the anomalous, non Ohmic, 
behavior observed experimentally in several multilayer systems. 
Our results show that this behavior can be explained in terms of the 
electronic structure of the slab only. No extra assumption for the
scattering at the interfaces is necessary. 
The connection of electronic structure and conductivity during layer-by-layer
growth is elucidated by analyzing the layer-projected conductivities.
\end{abstract}
\pacs{75.70.Pa,73.21.-b}
\Mmulti
\section{Introduction}
The understanding of the size-dependent conductivity in thin-film
multilayers is still an open question. The conductivity of various
magnetic/nonmagnetic multilayer systems 
\cite{bailey00,brueckl92,eckl94,faehler98,urbaniak98,faehler00}
has been measured as a function of the thickness of the deposited
layer. These electrical measurements were made simultaneously by
in-situ conductance monitoring,
and the experimental data show characteristic features for all investigated
multilayers.
First of all, the conductance increases with increasing multilayer
thickness.
Second, the slope of the conductance increase is different for the
different metal layers, due to the corresponding residual resistivity of
the metals. Third, there is a characteristic drop in
the total conductance as soon as the magnetic layer is added on top of
the nonmagnetic layer. This behavior is quite general and was originally related
to additional scattering in the interface region due to intermixing
and disorder. 

Several models have been developed to understand the origin of this
behavior. Most of them have assumed free-electron behavior within the
constituent layers 
\cite{palasantzas97,harrison00,palasantzas00},
extending the Fuchs-Sondheimer approach 
\cite{fuchs38,sondheimer52}.
Disorder at the interface was introduced. A quantum mechanical
description including surface scattering was presented by Fishman and
Calecki \cite{fishman89,fishman91} and by Trivedi and Ashcroft
\cite{trivedi88}.
A more realistic description of the multilayer was based on a tight
binding model including disorder at the interface \cite{bailey00}.
Ab initio electronic structure calculations of magnetic multilayers
have been performed by several groups
\cite{schep95,zahn95,butler96,zahn98,blaas99a}.
It was shown that the electronic structure of the multilayer plays a
very important role for the understanding of the transport properties
and giant magnetoresistance \cite{binasch89,baibich88}.
None of the calculations were performed for the in-situ conductance-monitoring 
considering the incremental conductance contributed by each
atomic layer. The aim of this work is the understanding of the size dependence 
and the
evolution of the Co/Cu multilayer conductivity
during the growth, starting from ab-initio electronic structure
calculations.

\section{Computational approach}
The calculations are based on Density Functional Theory (DFT) in the Local 
Density Approximation (LDA) and 
a recently implemented version of a Screened Korringa-Kohn-Rostoker
Green's Function method 
\cite{zahn98,anderson92,szunyogh94,zeller95,wildberger97}.
For layered systems the advantage of this transformation is the
N-scaling behavior (with N the number of layers in the system)
which enables us to treat systems with a large number of atoms in
the unit cell and arbitrary two-dimensional 
or three-dimensional cell periodicity.
The potentials were assumed to be
spherically symmetric inside the Wigner-Seitz sphere as in the Atomic Sphere
Approximation (ASA) but a multipole expansion of the charge density
has been taken into account up to  
$l_{max}=6$ (angular momenta up to $l_{max}=3$ have been used for the
wave functions). 
For the exchange correlation functionals the expressions given by
Vosko, Wilk and Nusair \cite{vosko80} were used.
Brillouin zone (BZ) integrations have been performed by means of special 
points methods \cite{monkhorst76}.
To simulate the in situ conductance-monitoring of the Co/Cu
multilayer during growth we model  
the system by the following assumptions. We consider a monolayer
growth along the (001) direction.
Co and Cu are assumed to occupy an ideal fcc lattice with a lattice constant
of
6.76 a.u., without any
lattice mismatch or distortion at  
the interfaces and surfaces, while no intermixing at the interfaces is 
allowed.

In the experimental setup an insulator or semiconductor buffer or target
material is used, this is 
simulated by vacuum in our calculation. That means a free standing slab with 2D periodicity in
plane is considered. For the calculations we
started from a Cu slab of 2 monolayers (ML) thickness. Cu was added
layer by layer until a thickness 
of 5 ML which corresponds to about $9\AA$. We continue by
adding Co up to the same thickness. 
Thus the cycle was repeated. 
The short-hand notation for the considered systems is
$Cu_{n_1}/Co_{n_2}/Cu_{n_3} ...$ denoting the individual layer
thicknesses in monolayers starting from the bottom Cu layer 
. 
Every multilayer corresponds to an experimental film after  
deposition of a complete ideally flat atomic layer. The electronic
structure was determined  self-consistently for each system under
consideration. 
Concerning the magnetic configuration of the mutlilayer slabs we assumed
parallel alignment of the Co layer magnetization which is
energetically preferred in Co/Cu multilayers 
\cite{lang96a} with 5 ML Cu thickness.

\section{Transport theory}
To calculate the electrical conductivity of the considered layered
structures we use the semiclassical Boltzmann theory in
relaxation time approximation \cite{zahn95,mertig99}. 
A spin-independent isotropic
relaxation time $\tau$ is assumed. This is of course a rough
approximation which concentrates on the  
anisotropic electronic structure of the system but, as our results show, it is
sufficient for the explanation of the effect. Moreover, considering the
anisotropic scattering would change our 
results only quantitatively, and not qualitatively and thus it serves
our current purpose of understanding 
the conductivity of thin
multilayers based on the electronic structure of ideal systems. Neglecting spin-flip scattering 
Mott's two current model can be used \cite{mott64}. With these simplifications 
the in-plane conductivity $\sigma$ is given by a Fermi
surface integral over the in-plane component of the Fermi velocity
$v_\VECk^s$ squared \cite{mertig99}
\begin{equation}
\sigma = \frac{e^2}{4\pi^2\hbar} \tau \sum_s \int \DTK 
\delta(E_\VECk^s - \ef) {v_\VECk^s}^2 .
\label{eqsig}
\end{equation}
$E_\VECk^s$ are the energy eigenvalues and $\ef$ is the Fermi level. 
s denotes the spin directions.
The k-integration is performed over the two-dimensional Brillouin
zone of the slab. The Fermi 
surface reduces to a set of closed lines
in the plane perpendicular to the growth direction 
.
The result of Eq.\ (\ref{eqsig}) is a slab 
conductivity per unit cell for a two-dimensional system. If we consider only one
monolayer then we obtain a conductivity per atom which compares
directly to the bulk conductivity.  

To analyse the spatial distribution of the current a local
conductivity is introduced by projecting the conductivity on each
atomic layer $i$
\begin{equation}
 \sigma_i^s =  \frac{e^2}{4\pi^2\hbar} \tau 
\int \DTK 
 \delta(E_\VECk^s - \ef) {v_\VECk^s}^2
 P^s_\VECk(i)
 \label{eqlocalsig}
\end{equation}
where 
\begin{equation}
 P^s_\VECk(i) = \int_{ASA} \DTR \left| \Psi_\VECk^s ({\mathbf R_i + \mathbf
r}) \right|^2  
 \label{eqprob}
\end{equation}
and $\Psi_\VECk^s ({\mathbf R_i + \mathbf r})$ are normalized wavefunctions of the 
slab. $\mathbf R_i$ are the basis vectors, denoting the different 
atomic layers, while the real space integration in Eq.\ (\ref{eqprob})
is restricted to the atomic sphere. 
Thus the total conductivity is just the sum over all the layers N, of
the slab and spin directions
\begin{equation}
\sigma = \sum_s \sum_{i=1}^N \sigma_i^s \qquad .
\label{eqsigma}
\end{equation}
\section{Ultrathin Cu film}
Following the ideas of Fuchs and Sondheimer
\cite{fuchs38,sondheimer52} the conductivity of a simple metallic film 
goes to zero with decreasing film thickness $D$, since in the limit of thin films: 
$\sigma \sim \kappa \ln(1/\kappa)$, where
$\kappa=D/l $, and $l $ denotes the mean free path. 
Considering no diffusive surface scattering we obtain the bulk conductivity.
In a realistic quantum mechanical calculation where full band structure effects
are present the situation is more complicated. 

\begin{figure}[htb]
\begin{center}
\begin{minipage}{\Mfactor\linewidth}
\includegraphics[width=\linewidth]{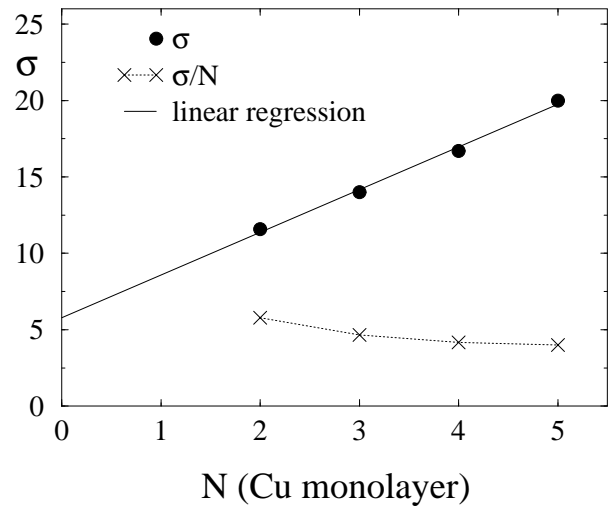}
\caption{
Film conductivity (dots) and conductivity per atom (crosses) as a
function of Cu layer thickness in arbitrary units.
} 
\label{fig_ultrathin}
\end{minipage}
\end{center}
\end{figure}  
In Fig.\ \ref{fig_ultrathin} we present  
the calculated conductivity of ultrathin Cu slabs with thicknesses
varying from two to five monolayers. 
The total slab conductivity increases almost linearly 
with the number of Cu monolayers. The conductivity per atom saturates to a
constant value after 4-5 ML. It is obvious that our
calculated conductivity does 
not extrapolate to zero as predicted by the Sondheimer model. This is
a direct consequence of the quantum mechanical treatment of the
surface. In the Sondheimer model the surface is described by the
classical electron distribution function. In the present calculation
the full quantum
mechanical Hamiltonian for the surface is used. 
As a result new types of eigenstates like
surface states could appear which are not included in the Sondheimer model. 
Due to the formation of surface states with a
high in-plane group velocity the conductivity per atom
(Fig.\ \ref{fig_ultrathin}) is increased for very
thin Cu slabs. It has to be mentioned that the constant relaxation
time that is used for all states at the Fermi level might be
inaccurate to obtain quantitative information.
\section{Conductivity of Cu/Co slabs}
\begin{figure}[htb]
\begin{center}
\begin{minipage}{\Mfactor\linewidth}
\includegraphics[width=\linewidth]{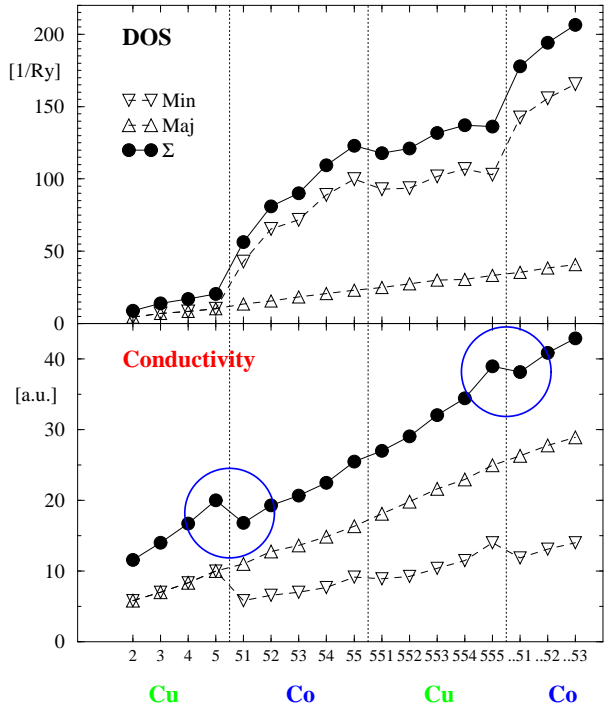}
\caption{
Evolution of
total and spin projected DOS [states/Ryd] at $\ef$ and 
total and spin projected conductivity [arbitrary units]
during Co/Cu multilayer growth.
} 
\label{fig_ntot}
\end{minipage}
\end{center}
\end{figure}  
Our calculated conductivity of the Cu/Co slabs is shown in
Fig.\ \ref{fig_ntot}, 
together with the DOS at the Fermi level. With our calculation we are able
to reproduce the experimentally observed behavior. Our results show
an increase 
with increasing slab thickness, while there is a drop in the slab
conductivity of about 20\% when the first Co monolayer is deposited on Cu. 
By adding more Co layers $\sigma$ increases further, and we observe no
significant effect 
when we start depositing Cu again. A drop is again seen when we repeat the 
Co deposition on top of Cu. The spin decomposition shows that for
majority electrons  
we have a monotonous increase in $\sigma $, while the peculiar,
non-Ohmic conductivity 
variation stems from the minority electrons. Here we point out that no
specific surface scattering mechanism is included in the
calculation. Moreover, a constant spin-independent relaxation time
was assumed, and the difference in the spin channels comes solely from the 
contrast of electronic structure in the two spin channels while differences 
in scattering which would require an anisotropic $\tau$ are ignored in this approximation.     

It is interesting to compare the evolution of the conductivity with the
changes of the slab DOS at the Fermi level, shown in Fig.\ \ref{fig_ntot}.  
For the majority channel we observe a linear increase with   
thickness since the local DOS at $\ef$ at the Co and Cu sites is very similar
for the majority channel and corresponds to the free electron like 
states above the filled 3d bands. 
The behavior of the minority DOS reflects the difference in the
electronic structure of the  
two materials. The DOS is much larger at the Co sites in comparison to
Cu, this causes the strong increase of the total DOS during the Co growth.  
The local DOS at the Co surface or 
at a Co/Cu interface is enhanced in comparison to the bulk Co, and
the total DOS is strongly increased by adding the
first two Co layers. Moreover, quantum confinement effects are fully included
in our calculation and cause the reduced increase when depositing the third Co
layer, and the small drop when the single Cu layer terminates the slab. 

Let us now concentrate on the drop of the conductivity when adding one
monolayer 
Co on the Cu film. This behavior is accompanied by a strong increase
in the total DOS since the Fermi level crosses the Co minority band. 
The spin projected conductivities demonstrate that the drop occurs in
the minority channel. 
By means of an averaged spin-dependent Fermi velocity
$\bar{v}^s$ 
\begin{equation}
\bar{v}^s =  
\left[
\int \DTK\delta(E_\VECk^s - \ef) {v_\VECk^s}^2
/n^s(\ef)
\right]^\frac{1}{2}
\label{eqvel_avg}
\end{equation}
one can rewrite Eq.\ (\ref{eqsig}) into 
\begin{equation}
\sigma =  \frac{e^2}{4\pi^2\hbar} \tau 
\sum_s   
n^s(\ef) \left.{\bar{v}^s}\right.^2 \qquad .
\label{eqsig_avg}
\end{equation}
$n^s(\ef) = \int \DTK\delta(E_\VECk^s-\ef) $ is the total DOS at Fermi level.
The expression in Eq.\ (\ref{eqsig_avg}) is useful to understand the
microscopic origin of the conductivity drop.
Following Eq.\ ({\ref{eqsig_avg}}) the reduction of the conductivity
is caused by a strong reduction of the averaged Fermi velocity in the
minority channel since the DOS is continously increasing.
For the
majority band we have a linear increase and no difference is seen when 
going from Cu to Co and vice versa since the Co d states are below the Fermi level. 
\begin{figure}[htb]
\begin{center}
\begin{minipage}{\Mfactor\linewidth}
\includegraphics[height=0.97\textheight]{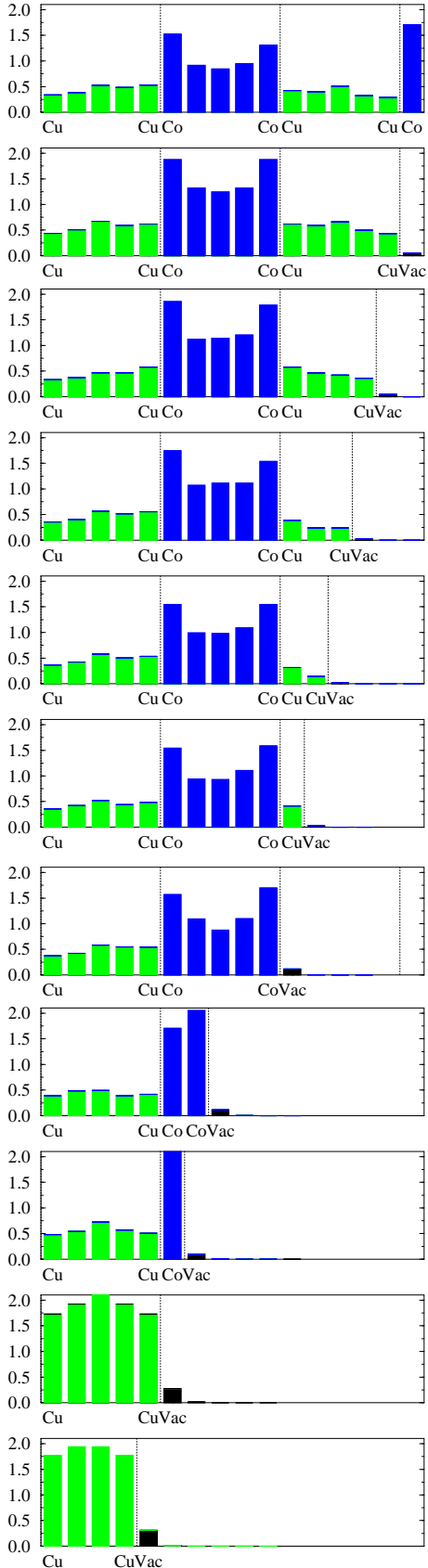}
\caption{
Evolution of the local conductivity of the minority channel 
(Eq.\ \ref{eqlocalsig}) 
[arbitrary units] during film growth.
} 
\label{fig_localsig}
\end{minipage}
\end{center}
\end{figure}  

To elucidate this Fermi velocity reduction the local conductivity of
the minority channel (Eq.\ \ref{eqlocalsig})
is shown in Fig.\ \ref{fig_localsig}, for nearly all slabs under
consideration.
For the pure Cu slabs all Cu layers contribute equally to the
conductivity, and we observe only a small 
reduction at the surface layers.  The situation changes drastically 
when we consider a single monolayer of Co on the Cu slab. As we can see
from Fig.\ \ref{fig_localsig} (third panel from bottom), 
the major contribution in the conductivity
comes from the Co layers, the Cu contributions however, are significantly
decreased by almost a factor of four in comparison to the Cu slabs
without Co coverage. According to Eq.\ (\ref{eqsig_avg}), this corresponds
to a reduction of the effective Fermi velocity by a factor of two and  
the total slab conductivity is decreased
(Fig.\ \ref{fig_ntot}). 

\begin{figure}[htb]
\begin{center}
\begin{minipage}{\Mfactor\linewidth}
\includegraphics[width=\linewidth]{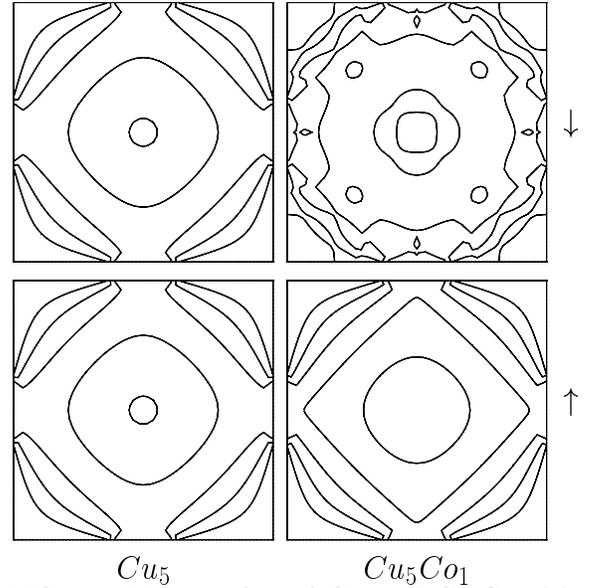}
\caption{ Fermi surface of $Cu_5$ and $Cu_5Co_1$ slabs for minority
(upper panel) and majority (lower panel) bands. Note that for 2d
slabs of finite thickness the Fermi surface reduces to a set of lines. 
} 
\label{fig_efsurf}
\end{minipage}
\end{center}
\end{figure}

This reduction of the Fermi velocity stems from a drastic change of the Fermi
surface topology when adding a single Co layer on top of Cu as shown
in Fig.\ \ref{fig_efsurf}. The deposition of
a simple Co layer destroys the free electron like Cu-slab Fermi surface in
the minority band while the majority Fermi surface is kept intact. 
As a result we end up with a complicated minority Fermi
surface dominated by Co d states of low velocity (Fig.\ \ref{fig_efsurf}).
Despite the low velocity of the Co d-states, Co layers have a high $\sigma$
because of the high DOS. Moreover, Co surface and interface states produce a
further enhancement.

The constant reduction of the Cu contribution implies also the
experimentally found 
proportionality of the conductivity drop in dependence on the Cu layer
thickness \cite{bailey00}.
The conducting states have a large probability amplitude at the
surface, which would give an amplification of the conductivity drop due to the enhanced
scattering cross section \cite{zahn97}.
This is in line with numerical results of Tsymbal \cite{bailey00}
assuming a larger potential disorder at the Co/Cu interface in
comparison to the bulk materials.

The addition of any extra Cu layer on top of the Co layer 
is accompanied by a similar conductivity contribution like in the
first Cu stack. 
In the minority channel
there is only a marginal decrease when we add 
one monolayer Cu on top of Co, this is compensated by the linear
increase in the majority 
channel, so that almost no effect is observed in the total $\sigma$. 
With increasing Cu layer thickness a symmetric profile
is formed (Fig.\ \ref{fig_localsig}, second panel from top). 
The total conductivity increases
monotonously.
The first Co layer of the second stack causes again a decrease of the
underlying Cu layer contributions and a drop of the total $\sigma$.
The combination of all details in the local conductivity discussed
above leads to the observed behavior as a function of the film
thickness (Fig.\ \ref{fig_ntot}), decreasing conductivity 
adding Co on Cu, and a linear increase when adding Cu on Co. 
The Cu thickness we consider is much smaller  
compared with the recent experiment of Bailey et al \cite{bailey00}. 
but as we can see from Figs.\ \ref{fig_ultrathin} and \ref{fig_ntot}, 
a constant conductivity per atom is established after 4-5 ML and larger 
thickness would only cause a monotonic increase in $\sigma$. As mentioned 
above our calculations are performed for a ferromagnetic configuration 
of the Co magnetization which is favored by the interlayer distance of 5 ML
of Cu.  The Cu thickness of $20\AA$ (11
ML) used in the experiment by Bailey et al. \cite{bailey99} and
Urbaniak et al. \cite{urbaniak98} corresponds to the second
antiferromagnetic maximum of interlayer exchange coupling in Co/Cu
and Py/Cu multilayers, respectively. However the actual magnetic configuration of the 
multilayer slab under consideration was not investigated in the experiments.

The microscopic picture discussed above could be applied to other systems with 
similar electronic structure like Py/Cu \cite{urbaniak98}, Fe/Ag
\cite{faehler98}, and Cr/Au \cite{brueckl92}.
Depending on the electronic structure of the particular system the
role of minority and majority electrons can be changed. The only
system that has to be checked in more detail is Cu/Ag
\cite{faehler00}. Cu and Ag are isovalent and the bandstructures are
quite similar. The conductivity drop in this system might be related
to alloying only.
\section{Conclusions}
Based on ab-initio calculations of the conductivity of Cu/Co slabs we have shown
that the variation during growth can be explained
by the changes of the electronic structure as a function of the
film thickness. In general the total conductivity of a multilayer slab
increases with increasing thickness. A pronounced conductivity drop is
obtained when the first Co layer is deposited on top of the Cu. Due to
interaction of the Co d-electrons with the Cu s-electrons the character of the 
minority Fermi surface is changed from sp to d-like. As a consequence the local
conductivity contributions of the Cu layers are reduced. Since the
total DOS at the Fermi level is increasing during film growth
the effect is related to reduced Fermi velocities. 
The microscopic
picture can be generalized to any multilayer consisting of noble and
transition metal layers. This effect is not in contradiction to the
conduction drop caused by interface scattering discussed by Bailey et
al. \cite{bailey99} but can be combined with it. Both effects exist
and interface scattering can even amplify the bandstructure effects.
\Mmultii

\begin{references}
\bibitem{bailey00}
W.E. Bailey, S.X. Wang, and E.Y. Tsymbal,
Phys. Rev. B \Vol{61}, 1330 (2000).
%
\bibitem{brueckl92}
H. Br\"uckl, PhD Thesis, Universit\"at Regensburg (1992),
%
\bibitem{eckl94}
H. Eckl, G. Reiss, H. Br\"uckl, and H. Hoffmann, J. Appl. Phys. \Vol{75}, 362 (1994).

%
\bibitem{faehler98}
S. F{\"a}hler, M. Weisheit, and H.-U. Krebs, Mat. Res. Soc. Proc. \Vol{502}, 139 (1998).
%
\bibitem{urbaniak98}
M. Urbaniak, T. Luci\'nski, and F. Stobiecki,
J. Mag. Mag. Mat. \Vol{190}, 187 (1998).
%
\bibitem{faehler00}
S. F{\"a}hler, M. Weisheit, K. Sturm, and H.-U. Krebs,
Appl. Surf. Sci. \Vol{154-155}, 419 (2000).
%
\bibitem{palasantzas97}
G. Palasantzas and J. Barna\'s,
Phys. Rev. B \Vol{56}, 7726 (1997).
%
\bibitem{harrison00}
W.A. Harrison, Phys. Rev. B \Vol{61}, 7766 (2000).
%
\bibitem{palasantzas00}
G. Palasantzas, Y.-P. Zhao, G.-C. Wang, T.-M. Lu, J. Barna\'s, and
J. Th. M. De Hosson,
Phys. Rev. B \Vol{61}, 11109 (2000).
%
\bibitem{fuchs38}K. Fuchs, Proc. Royal Soc. A \Vol{34}, 100 (1938).
%
\bibitem{sondheimer52}E. H. Sondheimer, Adv. Phys. \Vol{1}, 1 (1952).
%
\bibitem{fishman89}
G. Fishman and D. Calecki,
Phys. Rev. Lett. \Vol{62}, 1302 (1989).
%
\bibitem{fishman91}
G. Fishman and D. Calecki,
Phys. Rev. B \Vol{43}, 11581 (1991).
%
\bibitem{trivedi88}
N. Trivedi and N. W. Ashcroft,
Phys. Rev. B \Vol{38}, 12298 (1988).
%
\bibitem{schep95}
K.M. Schep, P.J. Kelly, and G.E.W. Bauer, Phys. Rev. Lett. \Vol{74}, 586 (1995).
%
\bibitem{zahn95}P. Zahn, I. Mertig, M. Richter, H. Eschrig,
Phys. Rev. Lett. \Vol{75}, 2996 (1995).
%
\bibitem{butler96}
W.H. Butler, X.-G. Zhang, D.M.C. Nicholson, T.C. Schulthess,
and J.M. MacLaren,
Phys. Rev. Lett. \Vol{76}, 3216 (1996).
%
\bibitem{zahn98}
P. Zahn, J. Binder, I. Mertig, R. Zeller, and P.H. Dederichs,
Phys. Rev. Lett. \Vol{80}, 4309 (1998).
%
\bibitem{blaas99a}
C. Blaas, P. Weinberger, L. Szunyogh, P.M. Levy, and C. Sommers,
Phys. Rev. B \Vol{60}, 492 (1999).
%
\bibitem{binasch89}
G. Binasch, P. Gr{\"u}nberg, F. Saurenbach, and W. Zinn,
Phys. Rev. B \Vol{39}, 4828 (1989).
%
\bibitem{baibich88}M.N. Baibich, J.M. Broto, A. Fert, F. Nguyen Van Dau,
F. Petroff, P. Etienne, G. Creuzet, A. Friederich, and J. Chazelas,
Phys. Rev. Lett. \Vol{61}, 2472 (1988)
%
\bibitem{anderson92}O.K. Anderson, A.V. Postnikov, and
S.Yu. Savrasov, Mat. Res. Soc. Proc. \Vol{253}, 37 (1992).
%
\bibitem{szunyogh94}
L. Szunyogh, B. \'{U}jfalussy, P. Weinberger,
and J. Koll\'{a}r, Phys. Rev. B \Vol{49}, 2721 (1994).
%
\bibitem{zeller95}
R. Zeller, P.H. Dederichs, B. \'{U}jfalussy, L. Szunyogh, and
P. Weinberger, Phys. Rev. B \Vol{52}, 8807 (1995).
%
\bibitem{wildberger97}
K. Wildberger, R. Zeller, and P.H. Dederichs, Phys. Rev. B \Vol{55}, 10074 (1997).
%
\bibitem{vosko80}S. H. Vosko, L. Wilk, and M. Nusair, Can. J. Phys. \Vol{58},
1200 (1980).
%
\bibitem{monkhorst76}H.J. Monkhorst and J.D. Pack, Phys. Rev. B \Vol{13}, 5188 (1976).
%
\bibitem{lang96a}
P. Lang, L. Nordstr{\"o}m, K. Wildberger, R. Zeller,
P.H. Dederichs, and T. Hoshino, Phys. Rev. B \Vol{53}, 9092 (1996).
%
\bibitem{mertig99}
I. Mertig, Rep. Prog. Phys. \Vol{62}, 237 (1999).
%
\bibitem{mott64}N.C. Mott, Adv. Phys. {\bf13}, 325 (1964)
%
\bibitem{zahn97}
P. Zahn, I. Mertig, R. Zeller, and P.H. Dederichs,
Mat. Res. Soc. Proc. \Vol{475}, 525 (1997).
%
\bibitem{bailey99}
W.E. Bailey, C. Fery, K. Yamada, and S.X. Wang,
J. Appl. Phys. \Vol{85}, 7345 (1999).
%
\end{references}
\end{document}